# The Never Ending Gale:
# its Role in Captain Robert F. Scott and
# his Companions' Deaths

Krzysztof Sienicki

*Chair of Theoretical Physics of Naturally Intelligent Systems*
*ul. Topolowa 19, 05-807 Podkowa Leśna, Poland, EU*
krissienicki@yahoo.com

(14 September 2011)

## Abstract

Polar historians and enthusiasts are aware that toward the end of March 1912, Captain Robert F. Scott reported in his journal a meteorological event, which was extraordinary as far as its length and strength was concerned. This event was the gale which according to Captain Scott, lasted nine/ten days. Were the laws of physics suspended at the end of March 1912 in the Antarctic? I have shown that the near surface winds in the Antarctic are self-organized critically and that the winds over the continent form an ergodic system. I have presented an analysis of wind events in the proximity of Captain Scott's camp and at Ross Island. By comparing wind events at these locations, and performing an analysis of a gale's wind duration and strength at One Ton Depôt, I concluded that Captain Scott's wind record was highly inaccurate. I concluded that the nine/ten day gale described by Captain Scott, that lasted from March 21 to 29, did not take place. This result combined with my previous analysis of Captain Scott's temperature record, shows that two black swan meteorological events: February 27-March 19, 1912 – Extreme Cold Snap and March 21-29, 1912 – Never Ending Gale reported by Captain Scott, did not take place. Therefore, I conclude that the deaths of Scott, Wilson and Bowers were a matter of choice rather than chance.

**Keywords:** Antarctic, Exploration, Weather, Self-organized Criticality, Captain Scott, Robert F. Scott, Amundsen, Ross Ice Shelf

## 1. Introduction

During the planning and executing stages, the Captain Scott and Roald Amundsen expeditions were definite events which were reduced into numbers: miles per day, calories per day, temperature, efficiency, work, forecast, friction, etc. Both the explorers and the members of their expeditions, in different degrees of proficiency and expertise, transformed these numbers into their everyday lifestyles and actions. Roald Amundsen and Captain Scott had accumulated knowledge and experience in Polar Regions while planning the South Pole expedition. Amundsen gained his skills during the *Belgica Expedition* (1897–99)[1] and Northwest Passage (1903–1906).[2] Captain Scott acquired his familiarity with the Antarctic throughout his *Discovery Expedition* (1901-1904).[3]

One of the most perplexing questions concerns what actually took place during the final days of the struggle of Captain Scott and his companions to the Cape Evans home base *en route* from the South Pole.

Life in Antarctica is inevitably related and dictated by the modes, severity, and apparent unpredictability of the weather. Then and now, a small weather window of opportunity appears for humans to explore the vast Antarctic continent. The size of this opportunity is simply measured by air temperature, human endurance and mileage to be crossed. If the distance travelled was to be extended, it would mean that endurance and organization would have to be better, meaning, better performance on a personal and organizational level would be needed. The game to reach the South Pole was a performance game. We know the results: Amundsen first, Scott second. We know or at least assume to know that Roald Amundsen's trip to the South Pole was a flawless effort, whereas Captain Scott suffered on many occasions.

However, as far as the Roald Amundsen and Captain Scott expeditions and reaching the South Pole is concerned, it is useful to make a *ceteris paribus* assumption. The assumption is that, in spite of different methods, means, and human effort, both expeditions were able to reach the South Pole and both teams were capable of returning safely to the base camp at Framheim and Cape Evans (Hut Point), respectively. This leaves the weather, understood as a combination of temperature and wind velocity, as the only *independent* variable.

I just mention two related examples. The first is Roald Amundsen's premature attempt to reach the South Pole on 8 September 1911, which had to be abandoned due to extremely low temperatures. The extremities of these temperatures are not due to any *absolute* temperature values, but because the dogs could not withstand the impact.[4] The second example is the four day blizzard, which struck Captain Scott's South Pole party at the foot of the Beardmore Glacier on December 5, 1911.

After examining the weather record of both the Roald Amundsen South Pole party and the Scott South Pole party, a list can be presented of temperature and wind velocity records measured every day of their trip. Just by looking at the respective plots of temperature and/or wind velocity versus time, rather regular (expected) variations of these variables can certainly be observed. Even the just mentioned four-day-blizzard would fall into this ordinary category. Only two meteorological events described by Captain Scott would fall into the category of *extremely rare* weather events, as observed (rouge wave) by Ernest Shackleton just before landing on the south shore of the South Georgia Island.[5] These weather events are sometimes called in power-laws jargon[6]: a *black swan*, *outliers*, *dragon-kings* or rogue wave (solitary wave).

Two black swan events were described by Captain Scott, and one event followed the other. The first black swan event occurred on a February 27-March 19, 1912 – Extreme Cold Snap, and the second occurred on a March 21-29, 1912 – Never Ending Gale.

Captain Scott and Roald Amundsen expeditions may be digitalized and analyzed, as in modern terms, we are accustom to think the entire Universe is a digital machine.[7]. Such an observation may be quite palpable for anyone who has ever planned a vacation trip or even a more challenging self-sustained expedition. Indeed, the Antarctic Manual[8] issued by the Royal Geographical Society (RGS) was intended as a digital description of various aspects of the planned Captain Scott *Discovery Expedition*. Unfortunately, instead of promoting "correct digits" the Antarctic Manual was a fairly random collection of different papers on various subjects related to the expedition. The RGS Secretary, Sir Markham was straightforward about his input[9]

"I planned an Antarctic Manual on the lines of the Arctic Manuals prepared for the expedition of 1875-76, securing the services of Mr. G. Murray as editor. It proved very useful, the first part containing instructions and information by leading men of science, and the second part being the narratives of Biscoe, Balleny, Dumont d'Urville, and Wilkes, with papers on polar travelling by Sir Leopold M'Clintock and on the exploration of Antarctic lands by Arctowski."

More importantly, the Manual was used by Sir Cements Markham to promote his interests by selecting certain contributors, rather than a sound scientific and digital handbook. In particular, an article by Sir Leopold McClintock, reprinted in the manual, presented *incorrect values* of sledging dog food rations. The rations advised by McClintock were half of what they should be. This error of negligence by McClintock, Markham, Murray, Scott and Shackleton pushed the whole history of British Antarctic exploration into the *Heroic* Age of Antarctic Exploration instead of into the *Grand* Age of Antarctic Exploration. See Appendix for more details.

It seems that Roald Amundsen and his team were sufficiently skilled at keeping and following the right numbers. Amundsen, unlike Captain Scott, was able to capture the right digits and the relationships between them. He followed what I call, 'fine tuned polar exploration' which includes subtle digital values and their relationships between nature (physical and life) and human needs and actions.[10]

Although the Amundsen and Captain Scott expeditions were preliminary logistic undertakings, this fundamental aspect of these expeditions was neglected by historians and bio- and hagiographers. Only a few exceptions exist scattered in professional journals. The majority of books concentrated on nostalgic lamenting over the suffering of Captain Scott and his companions. Authors close their eyes to scientific analysis while arguing that Captain Scott's expedition was a preliminary scientific expedition. The problem goes as far as data dragging, and/or attributing exaggerated data (increasing daily temperatures, wind velocities and its duration) to the Captain Scott expedition. In the case of Susan Solomon's work, data were even falsified in an attempt to prove her point.[11] However, falsifications or so-called 'minor adjustments', are not such a risky undertaking, as can be seen in Leonard Huxley's work. He got away with his falsification of Captain Scott's temperature records almost one hundred years ago.[12]

Polar historians and enthusiasts are aware that toward the end of March 1912, Captain Scott, this time without the support of Lt. Bowers or Dr. Wilson, reported in his journal a meteorological event which was extraordinary as far as its length and strength was concerned. This event was the *gale* which lasted nine/ten days. Captain Scott's entry tells us, "Thursday, March 29. – Since the 21st we have had a continuous gale from W.S.W. and S.W." and continues "Every day we have been ready to start for our depot 11 miles away, but outside the door of the tent it remains a scene of whirling drift."[13]

A careful reader can note from Captain Scott's own descriptions or from Simpson's meteorological record and analysis, that frequently observed gales (or blizzards) at the Ross Ice Shelf last about one-two days. One or two days are to be expected, but nine or ten days?

Were the laws of physics suspended at the end of March 1912 in the Antarctic? Did it happen for a second time? Did it happen only at the actual location of Captain Scott's team and did it not occur at locations where simultaneous measurements were taken by other members of the *Terra Nova Expedition*?

## 2. Nature of Near Surface Winds in the Antarctic

I have already shown in my previous work[14] that the first black swan[15] in the meteorological record (February 27-March 19, 1912 – Extreme Cold Snap) of Captain Scott did not happen, and the laws of physics were not suspended on the Ross Ice Shelf.

Captain Scott posed some questions which also concern the laws of physics. At the entrance to the Beardmore Glacier, Captain Scott's South Pole party was stopped and held in their tents for four long days by a blizzard or gale as he called it in the *Message to the Public*. Captain Scott posited the question, "What on earth does such weather mean at this time of year?" and wonders: "Is there some widespread atmospheric disturbance which will be felt everywhere in this region as a bad season, or are we merely the victims of exceptional local conditions? If the latter, there is food for thought in picturing our small party struggling against adversity in one place whilst others go smilingly forward in the sunshine."[16]

Captain Scott's reasoning is interesting indeed. It may sound as a beacon to his lines in his famed *Message to the Public* "our wreck is certainly due to this sudden advent of severe weather which does not seem to have any satisfactory cause."[17] But in the above comments by Capitan Scott two variables: distance and time, are missing. The distance between the entrance to the Beardmore Glacier and McMurdo Sound is about 600 km. It is more than the distance between London and Edinburgh. Does anyone expect or anticipate analogous weather conditions between these two cities? Cutting that distance in half one finds himself in about the position of (Leeds/Manchester) the One Ton Depôt or Schwerdtfeger (-79.904°, 169.97°) automated weather station, and Captain Scott's last camp. This was the very place where the second black swan meteorological event struck at gale strength and held the party in their tent for nine/ten days. Is it really possible that such a gigantic (duration and strength) wind event can actually take place? Is it possible that such a gigantic wind event can take place as geographically isolated phenomenon?

The question of nature and the wind dynamics over the Antarctic continent has been a central issue since the early developments of global earth atmospheric air circulation.[18] Four Americans, Charles Tracy (1843), William Ferrel (1856/1859), Matthew F. Maury (1859) James H. Coffin (1875), at around the same time, independently arrived at a similar description of global air circulation.[19] However, it may be of an even greater surprised to learn that American explorer Charles Wilkes, who sailed into the Antarctic Ocean, also published a book in which he correctly accounts for global circulation. In his book, "Theory of the Wind" published in 1856, Captain Wilkes explains[20]:

> It is known that the earth revolves on its axis, in a direction from west to east, increasing from the Poles to the Equator, where it attains a velocity of 1000 miles per hour, - so that the air, in passing from the highest latitudes towards the Equator, progressively arrives at regions of increased rotary velocity; and, as they cannot keep pace with this increase of motion, they necessarily hang back, and form currents flowing in a direction opposite to that of the rotation of the earth, or from east to west, and thus, by these combined efforts, the northern and southern currents of air are deflected and modified, so as to become the permanent north-easterly and south-easterly currents, forming the magnificent phenomenon of the Trade Winds.
> The theory is summed up, that whenever the air has greater velocity of rotation than the surface of the earth, a wind more or less westerly is produced, and when it has less velocity, a wind having an easterly tendency results.

When investigations were first made, many authors assumed that katabatic winds are downslope gradient-driven flows.[21] The long-wave radiative loss to space, leads to the build-up of high density cold air over the Antarctic dome. The presence of dome slope (relative height) induces a horizontal temperature gradient, leading to a downslope horizontal pressure-gradient katabatic force. Katabatic winds are then driven toward the coastal fringes and outwards towards the Antarctic Ocean. This scenario may be pertinent during the winter time. However, during the equinoctial seasons, katabatic winds are also observed. Such an observance takes place despite the simple fact that radiative loss, mentioned above, does not takes place. On the contrary, the Antarctic dome surface gains energy. While using loosely defined parameters, various rationalizations have been developed.

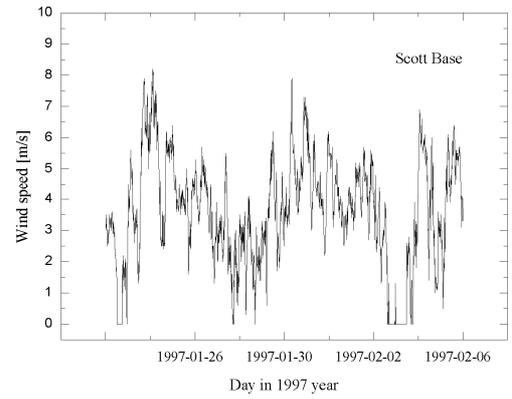

Figure 1. An example of a near surface wind event (wind velocity $v(t)$) measured at the Antarctica New Zealand Scott Base.

These rationalizations account for the expected different dynamical behaviour at *meso*- to synoptic-scale regimes of driving forces of near surface wind in the Antarctic.[22]

However, even a cursory examination will show that the vortex over the Antarctic continent is slowly-driven by a solar radiation non-equilibrium system with many degrees of freedom and a high level of nonlinearity. Figure 1 depicts a near surface wind velocity $v(t)$ measured at the Antarctica New Zealand Scott Base station located at the southern tip of Hut Point Peninsula of Ross Island. The erratic behaviour of this time series is clearly visible.

In equilibrium statistical mechanics, the inverse of temperature $(1/k_B T)$ is a constant physical system parameter for a given state with energy $E$. However; all natural phenomena exhibit significant spatiotemporal temperature fluctuations (*meso*scale). For this reason, such phenomena must be regarded as non-equilibrium systems.[23]

In order to illustrate this behaviour, I analyzed the event-like structure of wind events which are partially depicted on Fig. 1. Their shown time scale can be divided into two compartmental events directly related to the size and the duration of the events. The first case of non-zero measurements of the wind velocity I will call the wind event size $(w_s)$. I define the wind event as $w_s = \int v(t)dt \approx \sum v(t)\Delta t$ for *successive* non-zero wind velocities, where $\Delta t$ is the size of the measurement bin. In the second case, the second compartment can be considered duration of wind event $(w_t)$ and duration of quiescent wind event $(w_q)$, respectively.

To analyze wind behaviour, I will use its size distribution and occurrence frequency distribution. When estimating a cumulative *size* distribution $p_s(w_s)$, wind *duration* distribution $p_t(w_t)$, and wind *quiescent* distribution $p_q(w_q)$ for a given data set, the number of wind events of the size $w_{s,t,q} \geq 0$ or greater is counted. The following power-law cumulative size distributions is used to fit measured wind velocity

$$p_s(w_s) \propto w_s^{-\alpha_s}, p_t(w_t) \propto w_t^{-\alpha_t} \text{ and } p_q(w_q) \propto w_q^{-\alpha_q}$$

By analysing wind velocity data from about 40 meteorological stations scattered across the Antarctic continent, I have found that data follow the above equations.(See Table 1) Thus, the wind events at these sites are self-organized criticality.[24] This is an important and fundamental property of the winds in the Antarctic.

The observed self-organized criticality of katabatic winds suggests that atmosphere over the Antarctic continent is organized into many interacting katabatic air cells forming the Polar Cell – circumpolar near surface wind regime. The motion of these katabatic cells is driven by local rather than continental forces, predominantly by a downslope buoyancy forcing. Each air cell contributes to the katabatic air cell at a certain threshold. Hence, the air over the Antarctic Plateau piles up on ice sheet-formed gullies and valleys while relaxing its excess along contoured orography of coastal fringes or glacier slopes. At a given time, near surface air over the Antarctic continent contains a great number of volatile air cells dissociated by close to equilibrium relaxed air cells, or the opposite.

Volatile and relaxed air cells are transient. These cells are formed much faster than the slow time-scale of driving force resulting from incoming solar radiation and the long-wave radiative loss to space. This process means the air system evolves through a sequence of states that are infinitesimally close to equilibrium in a cycle: driving, katabatic event, and relaxation. Transition between volatile air cells occurs through katabatic wind events which restore and redistribute driving field forces. The katabatic wind events, duration, and quiescent wind event have no preferred scale, and their sizes and durations follow power-law distributions. The numerical values of the exponents depend on the geographical location measured by the slope of the Antarctic dome. The wind field over the Antarctic is likely to be of a critical state. These are the essential factors of the mean field behaviour of near surface katabatic winds in the Antarctic or over arbitrary ice dome on Earth or on other planets.

Table 1. Selected locations with respective scaling parameters.

| Location /Scaling | $\alpha_s$ | $\alpha_t$ | $\alpha_q$ |
|---|---|---|---|
| The Antarctic Continent (AC) | $1.06 \pm 0.01$ | $1.49 \pm 0.04$ | $2.26 \pm 0.05$ |
| McMurdo (McM) | $1.31 \pm 0.07$ | $2.27 \pm 0.08$ | $4.77 \pm 0.16$ |
| Scott Base (SB) | $1.13 \pm 0.01$ | $1.39 \pm 0.03$ | $2.68 \pm 0.06$ |
| Schwerdtfeger (SCH) | $1.27 \pm 0.01$ | $1.39 \pm 0.03$ | $2.46 \pm 0.04$ |

Recently Magdziarz and Weron[25] derived a version of the Khinchin theorem for Lévy flights. Their result gives us a tool to study ergodic properties of Lévy $\alpha$ − stable distributions including the above reported power-law distributions of wind events in the Antarctic. The result of combined cumulative distribution functions of all wind events for the entire Antarctic Continent is given in Table 1. This confirms that wind over the Antarctic reaches a state of self-organized criticality and also that wind circulation shows *ergodic* behaviour. Therefore, a fractional Fokker-Planck equation for the probability distribution of air particles whose motion is governed by a non-Gaussian Lévy-stable noise, would be a suitable theoretical approach in the description of near surface winds over the Antarctic continent.

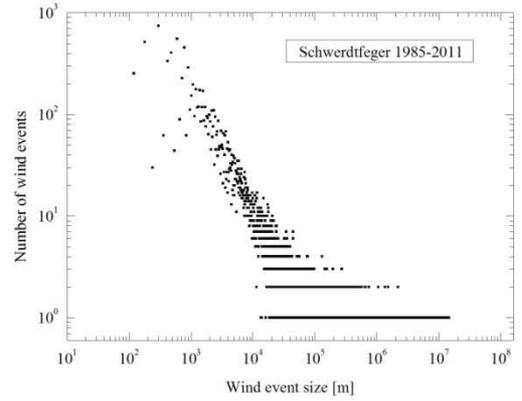

Figure 2. Cumulative wind size distribution function at Schwerdtfeger weather station located near One Ton Depôt and the final camp of Captain Scott's party.

## 3. Results and Discussion

In my analysis, I used 26 years of wind data measured at Schwerdtfeger and at the Antarctica New Zealand Scott Base[26] station located at the southern tip of Hut Point Peninsula of Ross Island, Antarctica. New Zealand's Scott Base is directly "facing" all meteorological events at the Ross Ice Shelf.

Both, Scott Base and Schwerdtfeger stations are not exceptions to the self-organized criticality of wind events shown in the above preceding sections. In Figure 2, I have depicted the cumulative size distribution function $p_s(w_s)$ for "raw" data for Schwerdtfeger. Wind event size distribution at this station shows typical power-law behaviour with scaling parameter $\alpha_s = 1.27 \pm 0.01$. Similar wind size scaling parameters calculated for Scott Base and McMurdo stations are $\alpha_s = 1.13 \pm 0.01$ and $\alpha_s = 1.31 \pm 0.07$, respectively. See Table 1.

Up to now, the belief that arbitrary size wind event may happen in the Antarctic was taken for granted by polar enthusiasts and authors. Although the nine/ten day gale at the end of March 1912, described by Captain Scott, was indeed an extraordinary event, no one bothered to look at meteorological data.

I have shown above, that due to the evident scaling property of wind event size at Schwerdtfeger and the remaining stations/locations, it can be argued that at this particular location, which is very close to the last camp of Capitan Scott's party, *arbitrary* wind event size is certainly possible. Thus, one can attempt to make a scholarly inference based on the above analysis of the scaling properties of wind events, suggesting the nine/ten days gale reported by Captain Scott did indeed take place.

However, such a conjuncture is not correct. The occurrence of arbitrary wind event size is only a theoretical possibility that a power-law relationship $p_s(w_s) \propto w_s^{-1.27}$ holds at Schwerdtfeger station, without physical limits or underlying physics laws.

On a grander scale than the Schwerdtfeger location, I have shown above that wind reaches a state of self-organized criticality over the whole Antarctic continent. The movement of the wind takes place within the polar cell boundary and the transported pool (mass) of air is limited

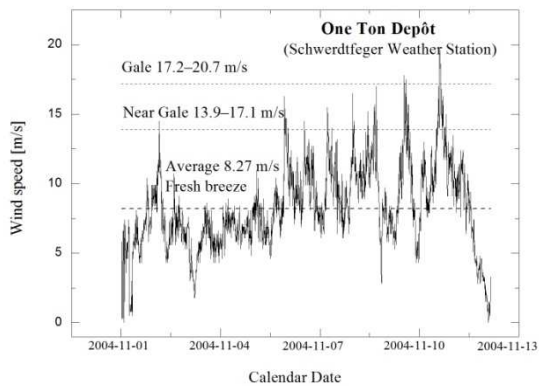

Figure 3. The longest and strongest wind event occurred at Schwerdtfeger station (One Ton Depôt) between Nov. 1 and Nov. 13, 2004.

and must be conserved. Therefore, no arbitrary wind event size can be observed.

Another important and limiting factor is the time needed to pass, in order for the observer to note arbitrary small or big event. And the third factor is that wind events at the Ross Ice Shelf are correlated. This means that whatever size wind event occurs at, for example, the Schwerdtfeger location, a similar in size (though delayed in time) event occurs at the Scott Base and McMurdo station. Alike correlations were also found between near surface air temperatures at these locations.

Capitan Scott reported that on March 21-29, 1912, a blizzard of gale force confined the party in the tent. It is fair to assume that Captain Scott, as a Royal Navy officer, was well trained and acquainted in describing wind force according to the Beaufort scale.[27] In this empirical measure, wind velocity in the range of 17.2–20.7 m/s (≈62–74 km/h) is named a *gale* – Beaufort number 8. Captain Scott used the term gale or moderate gale 59 times in his journal. Whenever, it was possible I checked Captain Scott's description against the *Terra Nova* or Simpson record and I confirmed that Captain Scott's description was accurate, with only one exception: the gale of March 21-29, 1912.

The Beaufort scale was a standard for ship log entries on Royal Navy vessels in the late 1830s. The Beaufort scale is rather a phenomenological scale.[28] The biggest pitfall of this scale is the lack of a time frame at which a given wind velocity has to be sustained in order to fall into a given category. Wind gusts and continuous fluctuations additionally complicate the issue.

Nevertheless, I have assumed that if within a 24h time frame, at least a *onetime* recorded wind velocity was equal or higher than the lower bound gale velocity (17.2 m/s ≈62km/h), and then a 24h frame was called "gale-day". If two gale-days were recorded in two, three, … *consecutive* days then such wind event is called a two day gale, three day gale, *etc*. This assumption is truly a very moderate one as a single wind gust would make a gale-day.

The general behaviour and structure of wind events have already been presented in Fig.1 for New Zealand's Scott Base. In spite of the erratic behaviour of wind events, a careful examination led me to conclude that self-organized criticality of wind events is observed across the Antarctic continent. Due to the above mentioned scaling properties of wind events, some assumptions can be made about the longest possible wind event at the location of Captain Scott's last camp in the proximity of One Ton Depôt. The assumptions would then apply to events close to the modern Schwerdtfeger weather station. By taking into account the *whole* available record of wind data (1986-2011), I found that the longest and strongest wind event occurred between Nov. 1 and Nov. 13, 2004. The wind velocity structure of this event is depicted on Figure 3. The duration of a wind event is understood as a continuous record of wind velocity $v > 0$ m/s with a detection threshold of about 0.1 m/s.

The Nov. 1 - Nov. 13, 2004 wind event which lasted thirteen consecutive days, had a few important characteristics. First of all, there was an extremely fast wind escalation (rise).[29] For an observer from the ground, it appeared as a wind blast. The second important characteristic was the wave structure of the wind velocity changes. And the third was the sudden downfall of wind.

To illustrate Captain Scott's wind record, I have also shown the lower bound of the Beaufort scale for wind of gale, near gale, and fresh breeze strength on the same figure (Fig. 3). The implications from this comparison is rather palpable. Captain Scott claimed that at their final camp - about 11 miles from One Ton Depôt, his party encountered a nine/ten day long gale. It is self-evident from Fig. 3, that it is stretching the meaning to call this most severe recorded wind event, a gale. There are only two wind *gusts* (spikes) on Fig. 3 which hardly and briefly reach gale force. Lowering the Beaufort scale to near gale force on the same figure shows that the biggest wind event can hardly be called a near gale event.

The following definitions[30] may be useful to re-call how the Beaufort scale is related to land conditions:

*Fresh Breeze* — Branches of a moderate size move. Small trees in leaf begin to sway,
*Near Gale* — Whole trees in motion. Effort needed to walk against the wind,
*Gale* — Some twigs broken from trees. Progress on foot is seriously impeded.

In the Antarctic, the wind creates near surface drag force. This force drives along air diamond dust and sweeps into the air those snow grains which were deposited on the surface. Conditions called a blizzard are then created. Blizzard conditions are those in which visibility and contrast are severely reduced.

According to Captain Scott's last entry in his field journal on March 29, 1912, the combined conditions of a nine/ten day gale with blizzard conditions was to blame for their failure to reach the One Ton Depôt. In particular, Captain Scott's comments "Every day we have been ready to start for our depôt 11 miles away, but outside the door of the tent it remains a scene of whirling drift."[31]

The biggest wind event presented in Fig. 3 was rather a fresh breeze. Such a breeze is much below that of a gale. Thus, the biggest wind event recorded by a modern weather station, at the proximity of One Ton Depôt, was about two times smaller than the event reported by Captain Scott back in March, 1912, and it took place in *November* not in March.

In modern data, it is not difficult to find the biggest wind event at Schwerdtfeger weather station, in the month of March. Such an event was observed between March 24-31, 1994 and is depicted on Fig. 4. The presented wind observation in March 1994, further confirms that the nine/ten day gale described by Captain Scott was much greater in length and force than any wind observed on modern record. This particular wind event March 24-31, 1994 also reveals an already described feature of all wind events in the Antarctic. This characteristic is a sudden rise and downfall of wind separated by frequent wind humps. These wind humps are long time scale fluctuations of wind velocity. The definition of blizzard provided by the National Oceanic and Atmospheric Administration[32] requires *sustained* gusts of gale force or greater, and blowing (falling) snow which reduces visibility to less than ¼ mile.

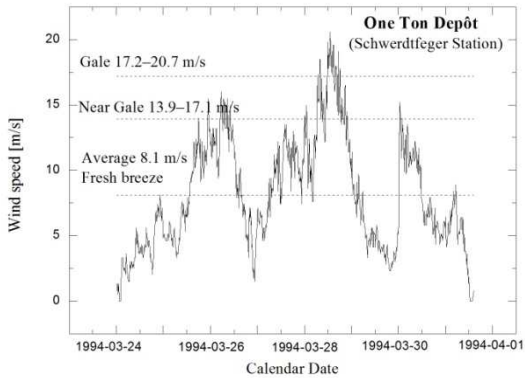

Figure 4. The biggest wind event at Schwerdtfeger weather station (One Ton Depôt), in the month of March. The event was recorded between March 24-31, 1994.

Returning to Fig. 3 and 4, it can be noted that within these extreme wind events one can select two different time windows: above the fresh breeze line and below the fresh breeze line. According to this definition, the conditions and the time "spend" by wind events below the fresh breeze line cannot be regarded as blizzard conditions. More importantly, when the weather conditions below the fresh breeze line are favourable, travel is thought possible although challenging. Looking at Fig. 4 it may be concluded that although this longest and strongest wind event lasted about 4 days (96h), there were encouraging weather conditions about half of the time. Thus, an average distance of about 22 miles could have been crossed by Captain Scott's sledging party. Obviously, the day & night routine would have to have been changed to a "pitch a tent and go when you can" schedule. Exactly the same observation can also be made by analyzing the biggest recorded November wind event at Schwerdtfeger station (One Ton Depôt). Therefore, the immobility of Captain Scott's party due to the nine/ten day gale from March 21-29, 1912 is dubious.

Let me investigate the next question of how many days a gale may last at the Ross Ice Shelf. A definition of gale-days was given above and a similar assumption is made in relation to weaker, near gale (13.9 m/s ≈50 km/h) wind events.

Figure 5 depicts the total number of consecutive gale-days recorded at the Schwerdtfeger and Scott Base weather stations, for every month of March from the whole 1985-2011 period. A wealth of knowledge is depicted in this figure. From my point of view and in relation to Captain Scott's record, it can be seen that in all the analyzed months of March (Figure 5, C & D), the longest consecutive gale and near gale winds lasted for three (3) and four (4) days. That is *far* below the black swan nine/ten day gale reported by Captain Scott. Figure 5 provides the second argument against Captain Scott's record of a nine/ten day long gale in late March 1912. The gigantic gale winds (and thus, the blizzard) reported by Captain Scott are of unfounded proportions and the laws of physics were not suspended in 1912.

Table 2. The weather Register of the Second Relief Party – Atkinson and Keohane.[33]

| Date | Time | Dry Bulb Temperature [°F] | Wind Force | Minimum Temperature [°F] |
|---|---|---|---|---|
| March 27 | 14:30 | +2.5 | 3 | - |
|  | 17:30 | -3.5 | 3 | - |
| March 28 | 7:00 | -6.5 | 2 | -6.0 |
|  | 12:30 | -15.5 | 3 | - |
|  | 17:00 | -6.5 | 1 | - |
| March 29 | 7:00 | -3.5 | 1-2 | -13.0 |
|  | 12:30 | -0.5 | - | - |
|  | 17:30 | -0.5 | 1-2 | - |
| March 30 | 7:00 | -8.5 | 1-2 | -16.0 |
|  | 12:00 | -3.5 | 3-4 | - |
|  | 17:30 | -5.5 | 3-4 | - |
| March 31 | 7:00 | -13.5 | 3-4 | -13.0 |

I presented the above two powerful arguments against Captain Scott's claim of an endless gale at the end of March 1912. Actually, the gale was not Captain Scott's first claim of unexpected weather. It is the second unfounded reason used by Captain Scott to explain "the storm which has fallen on us within 11 miles of the depôt at which we hoped to secure our final supplies."[34] But the truth was, the One Ton Depôt was not exactly salvation. The Depôt had not been stored, because the written orders of Captain Scott had not been followed. A salient mutiny was already taking place at Cape Evans in early 1912. Everyone had their own excuses for leaving the place. The dog driver, Cecil Meares, left on his own terms, G. Simpson the meteorologist, was apparently re-called for more (?) important duties in India, and Atkinson unwillingly assumed the command post without authority nor the means to enforce it. But it was Dr. Atkinson, who through his meteorological record of the rescue attempt in late March 1912, provides the final argument that Captain Scott forged his meteorological record. Not long after Cherry-Garrard and Demetri's premature dispatch of a relief attempt[35], a decision was taken to dispatch the Second Relief Party on March 26, 1912. Atkinson and Keohane started out alone from Hut Point to search for and help Scott's party. Neither of them could handle dog sledging and they resorted to the standard man-hauling technique. Their progress was very slow, only nine miles per day. The party travelled only to a point eight miles south of Corner Camp where Atkinson recorded: "At this date [March 30, KS] in my own mind, I was morally certain that the party had perished"[36]. Atkinson conjuncture is surprising, both because of its appeal to moral issues and its *post factum* character. An elementary calculation of the possible arrival time of Captain Scott's party, with a number of updates resulting from information provided by returning parties would give to Atkinson a date, March 30 (Corner Camp, ~40 miles (~64 km) from Hut Point)[37], which clearly questions his moral confidence.

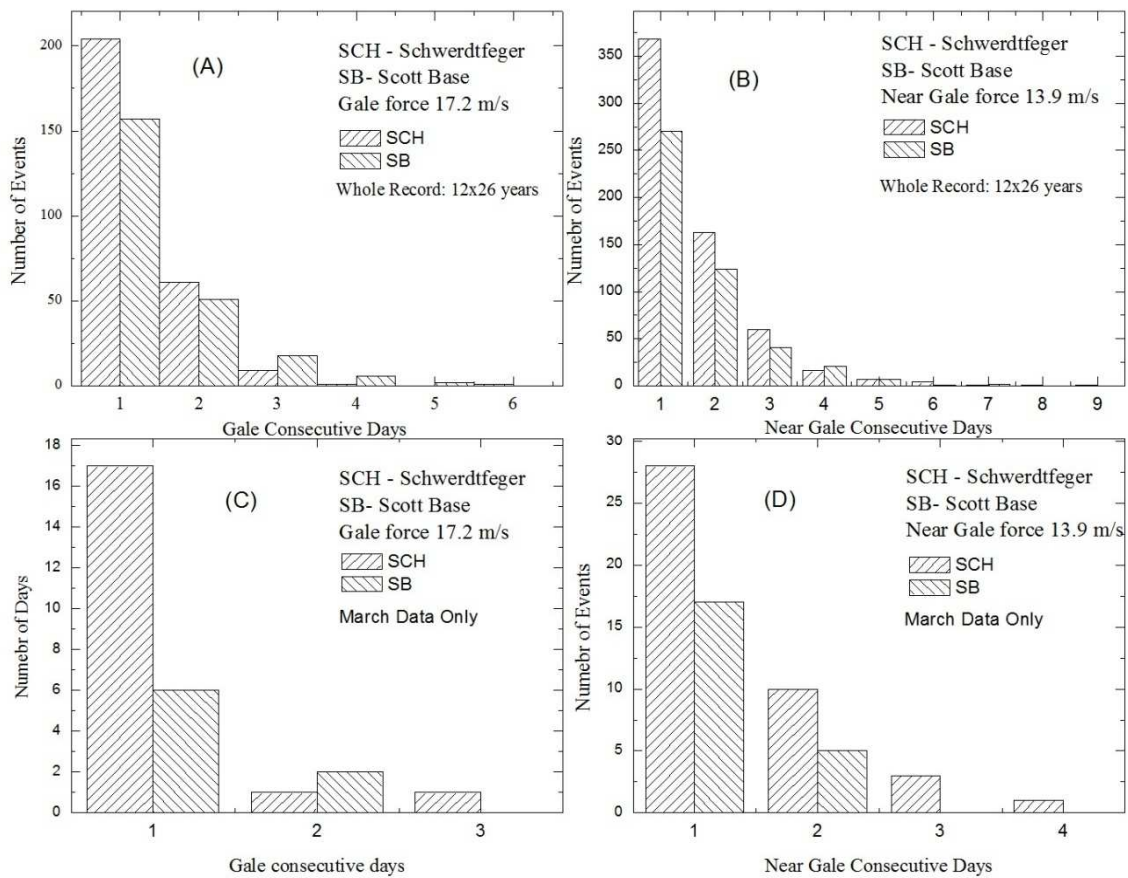

Figure 5. The total number of consecutive gale-days recorded at Schwerdtfeger (One Ton Depôt) and Scott Base weather stations, for whole record and for every month of March from the 1985-2011 period.

After leaving Hut Point, Atkinson and Keohane's party was fully exposed to air stream flowing northward along the Transantarctic Mountains.[38] Due to the already shown high correlations of temperatures and wind velocities along Captain Scott's route, including Atkinson and Keohane's course, both parties should have experienced the same or very similar weather conditions.

In the small Table 2, I have presented the weather Register of the Second Relief Party (Atkinson and Keohane). It is self-evident from this table, that the weather conditions recorded by Atkinson were *mild*. On March 26, Atkinson observed "The temperature was exceedingly low but the weather fair."[39] From March 27 until March 30, he recorded a wind force between 1 to 3. Thus, the strongest wind observed by the relief party was a gentle breeze; leaves and small twigs constantly moving according to the Beaufort scale description.

Comparing these conditions and the analysis of wind duration and strength as a raging gale at One Ton Depôt, the conclusion is that Captain Scott's wind record was highly inaccurate. The second black swan meteorological event did not take place.

There are two side issues which one may argue against my confirmation that Captain Scott invented the nine/ten day gale. These issues include the possibility that during his final days, Capitan Scott lost his ability for a reasonable, healthy judgement of reality due to his physical and mental state. There is no question that Captain Scott suffered greatly. There is no question that he was not sure about the actual date. Wilson, Bowers and possibly Captain Scott were uncertain whether the date was 21 or 22 March. However, even if we assume this 'one day of uncertainty' as a fact, it is obvious according to my analysis, that a *gale* that was one day shorter also never occurred.

Captain Scott wrote a number of farewell letters, a *Message to the Public* and a couple entries in his Journal during the last days of March 1912. All of these writings, their style, and finally the fact that they were written in his usual handwriting, confirm, at least up to this heartbreaking moment, that in spite of his words "It seems a pity, but I do not think I can write more", Captain Scott was able to reasonably and rationally judge reality.

## 4. Conclusions

In my research on Captain Scott's meteorological data reported by him and his party, I analysed two black swan events; February 27-March 19, 1912 – Extreme Cold Snap and March 21-29, 1912 – Never Ending Gale. In the analysis of the first black swan event I used a powerful computer-supported method of neural network simulation. This particular method enabled me to first of all, test the simulation's performance by using modern precise temperature data collected at locations associated with Captain Scott's expedition. Secondly, after multiple and satisfactory tests, the designed artificial neural network allowed me to use historic temperature data collected by Simpson, to reconstruct the true temperature record at the locations of Captain Scott's party in February and March 1912. Thus, I have shown that the temperature record of Captain Scott's party between Feb. 27-Mar. 19, 1912 was altered by Lt. Bowers and Captain Scott to inflate and dramatize the weather conditions.

In the present paper, I have used different methods to analyze what Captain Scott reported as a wind event of unprecedented proportions. The analysis of this second black swan meteorological event confirmed that despite my discovery of self-organized criticality of wind events in the Antarctic, a wind event (length and strength) described by Captain Scott did not take place.

Indeed these two black swan meteorological events followed each other. One could say, that in reality Captain Scott was struck by one dragon king[40] meteorological event starting on February 27 and continuing until March 29, 1912. Captain Scott's reported February 27-March 19, 1912 – Extreme Cold Snap and March 21-29, 1912 – Never Ending Gale were invented meteorological events and never occurred.

Therefore, I conclude that the deaths of Scott, Wilson and Bowers were a matter of choice rather than chance. It was a choice made long before the actual end of food, fuel and physical strength needed to reach the unsupplied by Meares, Simpson and Atkinson, salvation at One Ton Depôt.

## Acknowledgements

I wish to thank Professor Leszek Kułak from the Department of Theoretical Physics and Quantum Informatics at the Technical University of Gdańsk, Poland for assistance with neural network simulations. I also wish to thank him for the never ending discussions about entangled issues of the first principles of quantum mechanics. I am grateful to student Adrian Piasecki (M.Sc.) for his handling of weather data.

The author appreciates the support of the Antarctic Meteorological Research Center and the Automatic Weather Station Program for surface meteorological data, NSF (USA) grant numbers ANT-0636873, ARC-0713483, ANT-0838834, and/or ANT-0944018.

## Appendix: A Comment on Dog Use and Dog Rations.

## 'I Therefore Take the Liberty to Propose Three Cheers for the Dogs'
(October 24, 2008)

Lord Curzon's famous quotation cited in Bomann-Larsen[41] seems a fitting start for comments on Carl Murray's recent article in *Polar Record* entitled *The use and abuse of dogs on Scott's and Amundsen's South Pole expeditions* in which he presents a reassessment of the topic.[42] Murray devises a false presupposition that underlines all the puzzles it involves. Selecting from the original narratives only quotations that support his thesis, Murray claims that Scott applied utilitarianism to his expedition dogs in contrast to Amundsen who took a utilitarian approach by killing dogs and puppies. However, Murray disregards the same utilitarian killing (with Scott's euphemism '*reduce*' or '*remove*') of puppies during the Discovery and Terra Nova expeditions.[43] He also ignores what is self evident from Scott's *Discovery* diaries 'that the animals [dogs, KS] were carried on the sledges to keep them alive as fodder'.[44]

In a Table 1A, I have summarised maximum load and distance for sustainable polar sledging. The data were taken from

Table 1A: Estimated pooling power and distance of man and dog.

| McClintock Ref. 46 data for pooling power and distance. *The relationship between food weight for man and dog is unclear.* | | |
|---|---|---|
| 1 man | 1 load | 1 distance* /day |
| 1 dog | ½ load | ~ 2 distance/day |
| McClintock[45] & 1901(Ref.47) data for pooling power and distance. *Food weight for 2 dogs=1 man's food* | | |
| 1 man | 1 load | 1 distance/day |
| 2 dogs | 1 load | 1¼ distance/day |
| 2 dogs | 1 light load | 2 distance/day |
| Murray Ref. 42 data for pooling power and distance. *Food weight for 1 dog=1 man's food*. | | |
| 1 man | 1 load | 1 distance/day |
| 1 dog | 1 load | 1¼ distance/day |
| 1 dog | 1 light load | 1½ distance/day |

*McClintock's original estimation for distance was 13 miles and can be found only in the third edition of his narrative Ref. 46.

the third edition of the original McClintock narrative[46] and The Antarctic Manual.[47] Murray's rations[48] presented in Table 1A are erroneous and do not reflect historical evidence. Murray, by attributing his flawed data to Scott, is also incorrect.[49] The number of man or dog(s) and the distance are not *additive variables*.

From the table, one can notice that Scott's dog sledge, Southern Journey was doomed from the start. In calculations of dog rations, Scott and Wilson[50] used a half ration instead of the required 1 ration, because McClintock incorrectly suggested half.[51,52] Consequently, the dogs were underfed and soon after departure they 'seemed to lose all heart' and before long began to deteriorate. In order to advance his thesis, Murray, without evidence, attributes the sudden deterioration of dogs to 'tainted food.' The dogs may not have liked to be fed with dried codfish but the supplies were not tainted with a mysterious 'species of scurvy'. While on half rations, hard working sledge dogs were simply starving and soon all of them perished. In Scott's own words 'the real cruelty to a dog lies in over-working or underfeeding it'.[53]

One cannot restrict the notion of utilitarianism only to dogs, as in Murray's paper. The ponies of the *Terra Nova* expedition, in spite of Scott's previous pledge, miserably suffered to be shot, killed by Oates or Bowers with a pick-axe, killed by killer whales, to fall in a crevasse, or turned to hoosh and eaten by the main polar party at the foot of the Beardmore Glacier.[54]

Finally, I wish to remark that the treatment of draught animals by explorers was not unique or characteristic for their expeditions and/or individual characters [for example 55, 56, 57] The extraordinary physical and psychological burden of exploration was consequential to particular exploration spatiotemporal goals and the respective adding up for distance, time, speed, rations and endurance.